\documentclass[aps,prl,twocolumn,showpacs,superscriptaddress,groupedaddress]{revtex4}

\usepackage{graphicx}  
\usepackage{dcolumn}   
\usepackage{bm}        
\usepackage{amssymb}   
\usepackage[usenames]{color} 
\usepackage{amsmath}

\newcommand{\Btbw}      {\mbox{$\mathcal{B}(t\to Wb)$}}

\newcommand{\Gtbw}      {\mbox{$\Gamma(t\to Wb)$}}

\newcommand{\met}       {\mbox{\ensuremath{\slash\kern-.7emE_{T}}}}

\newcommand{\rargap}    {\mbox{ $\rightarrow$ }}

\newcommand{\ttbar}     {\mbox{$t\bar{t}$}}

\newcommand{\ppbar}     {\mbox{$p\bar{p}$}}

\begin{document}

\title{An improved determination of the width of the top quark}
\affiliation{Universidad de Buenos Aires, Buenos Aires, Argentina}
\affiliation{LAFEX, Centro Brasileiro de Pesquisas F{\'\i}sicas, Rio de Janeiro, Brazil}
\affiliation{Universidade do Estado do Rio de Janeiro, Rio de Janeiro, Brazil}
\affiliation{Universidade Federal do ABC, Santo Andr\'e, Brazil}
\affiliation{Instituto de F\'{\i}sica Te\'orica, Universidade Estadual Paulista, S\~ao Paulo, Brazil}
\affiliation{University of Science and Technology of China, Hefei, People's Republic of China}
\affiliation{Universidad de los Andes, Bogot\'{a}, Colombia}
\affiliation{Charles University, Faculty of Mathematics and Physics, Center for Particle Physics, Prague, Czech Republic}
\affiliation{Czech Technical University in Prague, Prague, Czech Republic}
\affiliation{Center for Particle Physics, Institute of Physics, Academy of Sciences of the Czech Republic, Prague, Czech Republic}
\affiliation{Universidad San Francisco de Quito, Quito, Ecuador}
\affiliation{LPC, Universit\'e Blaise Pascal, CNRS/IN2P3, Clermont, France}
\affiliation{LPSC, Universit\'e Joseph Fourier Grenoble 1, CNRS/IN2P3, Institut National Polytechnique de Grenoble, Grenoble, France}
\affiliation{CPPM, Aix-Marseille Universit\'e, CNRS/IN2P3, Marseille, France}
\affiliation{LAL, Universit\'e Paris-Sud, CNRS/IN2P3, Orsay, France}
\affiliation{LPNHE, Universit\'es Paris VI and VII, CNRS/IN2P3, Paris, France}
\affiliation{CEA, Irfu, SPP, Saclay, France}
\affiliation{IPHC, Universit\'e de Strasbourg, CNRS/IN2P3, Strasbourg, France}
\affiliation{IPNL, Universit\'e Lyon 1, CNRS/IN2P3, Villeurbanne, France and Universit\'e de Lyon, Lyon, France}
\affiliation{III. Physikalisches Institut A, RWTH Aachen University, Aachen, Germany}
\affiliation{Physikalisches Institut, Universit{\"a}t Freiburg, Freiburg, Germany}
\affiliation{II. Physikalisches Institut, Georg-August-Universit{\"a}t G\"ottingen, G\"ottingen, Germany}
\affiliation{Institut f{\"u}r Physik, Universit{\"a}t Mainz, Mainz, Germany}
\affiliation{Ludwig-Maximilians-Universit{\"a}t M{\"u}nchen, M{\"u}nchen, Germany}
\affiliation{Fachbereich Physik, Bergische Universit{\"a}t Wuppertal, Wuppertal, Germany}
\affiliation{Panjab University, Chandigarh, India}
\affiliation{Delhi University, Delhi, India}
\affiliation{Tata Institute of Fundamental Research, Mumbai, India}
\affiliation{University College Dublin, Dublin, Ireland}
\affiliation{Korea Detector Laboratory, Korea University, Seoul, Korea}
\affiliation{CINVESTAV, Mexico City, Mexico}
\affiliation{Nikhef, Science Park, Amsterdam, the Netherlands}
\affiliation{Radboud University Nijmegen, Nijmegen, the Netherlands and Nikhef, Science Park, Amsterdam, the Netherlands}
\affiliation{Joint Institute for Nuclear Research, Dubna, Russia}
\affiliation{Institute for Theoretical and Experimental Physics, Moscow, Russia}
\affiliation{Moscow State University, Moscow, Russia}
\affiliation{Institute for High Energy Physics, Protvino, Russia}
\affiliation{Petersburg Nuclear Physics Institute, St. Petersburg, Russia}
\affiliation{Instituci\'{o} Catalana de Recerca i Estudis Avan\c{c}ats (ICREA) and Institut de F\'{i}sica d'Altes Energies (IFAE), Barcelona, Spain}
\affiliation{Stockholm University, Stockholm and Uppsala University, Uppsala, Sweden}
\affiliation{Lancaster University, Lancaster LA1 4YB, United Kingdom}
\affiliation{Imperial College London, London SW7 2AZ, United Kingdom}
\affiliation{The University of Manchester, Manchester M13 9PL, United Kingdom}
\affiliation{University of Arizona, Tucson, Arizona 85721, USA}
\affiliation{University of California Riverside, Riverside, California 92521, USA}
\affiliation{Florida State University, Tallahassee, Florida 32306, USA}
\affiliation{Fermi National Accelerator Laboratory, Batavia, Illinois 60510, USA}
\affiliation{University of Illinois at Chicago, Chicago, Illinois 60607, USA}
\affiliation{Northern Illinois University, DeKalb, Illinois 60115, USA}
\affiliation{Northwestern University, Evanston, Illinois 60208, USA}
\affiliation{Indiana University, Bloomington, Indiana 47405, USA}
\affiliation{Purdue University Calumet, Hammond, Indiana 46323, USA}
\affiliation{University of Notre Dame, Notre Dame, Indiana 46556, USA}
\affiliation{Iowa State University, Ames, Iowa 50011, USA}
\affiliation{University of Kansas, Lawrence, Kansas 66045, USA}
\affiliation{Kansas State University, Manhattan, Kansas 66506, USA}
\affiliation{Louisiana Tech University, Ruston, Louisiana 71272, USA}
\affiliation{Boston University, Boston, Massachusetts 02215, USA}
\affiliation{Northeastern University, Boston, Massachusetts 02115, USA}
\affiliation{University of Michigan, Ann Arbor, Michigan 48109, USA}
\affiliation{Michigan State University, East Lansing, Michigan 48824, USA}
\affiliation{University of Mississippi, University, Mississippi 38677, USA}
\affiliation{University of Nebraska, Lincoln, Nebraska 68588, USA}
\affiliation{Rutgers University, Piscataway, New Jersey 08855, USA}
\affiliation{Princeton University, Princeton, New Jersey 08544, USA}
\affiliation{State University of New York, Buffalo, New York 14260, USA}
\affiliation{Columbia University, New York, New York 10027, USA}
\affiliation{University of Rochester, Rochester, New York 14627, USA}
\affiliation{State University of New York, Stony Brook, New York 11794, USA}
\affiliation{Brookhaven National Laboratory, Upton, New York 11973, USA}
\affiliation{Langston University, Langston, Oklahoma 73050, USA}
\affiliation{University of Oklahoma, Norman, Oklahoma 73019, USA}
\affiliation{Oklahoma State University, Stillwater, Oklahoma 74078, USA}
\affiliation{Brown University, Providence, Rhode Island 02912, USA}
\affiliation{University of Texas, Arlington, Texas 76019, USA}
\affiliation{Southern Methodist University, Dallas, Texas 75275, USA}
\affiliation{Rice University, Houston, Texas 77005, USA}
\affiliation{University of Virginia, Charlottesville, Virginia 22901, USA}
\affiliation{University of Washington, Seattle, Washington 98195, USA}
\author{V.M.~Abazov} \affiliation{Joint Institute for Nuclear Research, Dubna, Russia}
\author{B.~Abbott} \affiliation{University of Oklahoma, Norman, Oklahoma 73019, USA}
\author{B.S.~Acharya} \affiliation{Tata Institute of Fundamental Research, Mumbai, India}
\author{M.~Adams} \affiliation{University of Illinois at Chicago, Chicago, Illinois 60607, USA}
\author{T.~Adams} \affiliation{Florida State University, Tallahassee, Florida 32306, USA}
\author{G.D.~Alexeev} \affiliation{Joint Institute for Nuclear Research, Dubna, Russia}
\author{G.~Alkhazov} \affiliation{Petersburg Nuclear Physics Institute, St. Petersburg, Russia}
\author{A.~Alton$^{a}$} \affiliation{University of Michigan, Ann Arbor, Michigan 48109, USA}
\author{G.~Alverson} \affiliation{Northeastern University, Boston, Massachusetts 02115, USA}
\author{M.~Aoki} \affiliation{Fermi National Accelerator Laboratory, Batavia, Illinois 60510, USA}
% ...
%
%
\collaboration{The D0 Collaboration\footnote{with visitors from
$^{a}$Augustana College, Sioux Falls, SD, USA,
$^{b}$The University of Liverpool, Liverpool, UK,
$^{c}$UPIITA-IPN, Mexico City, Mexico,
$^{d}$ETH, Z\"urich, Switzerland,
$^{e}$DESY, Hamburg, Germany,
$^{f}$SLAC, Menlo Park, CA, USA,
$^{g}$University College London, London, UK,
$^{h}$Centro de Investigacion en Computacion - IPN, Mexico City, Mexico,
$^{i}$ECFM, Universidad Autonoma de Sinaloa, Culiac\'an, Mexico,
and 
$^{j}$Universit{\"a}t Bern, Bern, Switzerland.
$^{\ddag}$Deceased.
}} \noaffiliation
\vskip 0.25cm

\date{January 19, 2012}

\begin{abstract}

We present an improved determination of the total width of the top quark, $\Gamma_t$, 
using 5.4~fb$^{-1}$ of integrated luminosity collected by the D0 Collaboration 
at the Tevatron \ppbar\ Collider. The total width $\Gamma_t$ is extracted from the partial decay 
width $\Gamma(t\to Wb)$ and the branching fraction $\Btbw$. $\Gamma(t\to Wb)$ 
is obtained from the $t$-channel single top quark production cross section and
 $\Btbw$ is measured in $\ttbar$ events. 
For a top mass of $172.5\;\rm GeV$, the resulting width is 
$\Gamma_t = 2.00^{+0.47}_{-0.43}$~GeV. This 
translates to a top-quark lifetime of 
$\tau_t = (3.29^{+0.90}_{-0.63})\times10^{-25}$~s.
We also extract an improved direct limit on the CKM matrix element $0.81 < |V_{tb}| \leq 1$ at 95\% C.L.
and a limit of $|V_{tb'}| < 0.59$ for a high mass fourth generation bottom quark assuming unitarity of the fourth generation quark mixing matrix.
\end{abstract}

\pacs{14.65.Ha, 14.65.Jk, 12.15.Hh} 
\maketitle

\newpage

\section{Introduction}
\label{sec:intro}

The top quark is the heaviest known elementary particle and completes the quark sector of the  
standard model (SM). It differs from the other quarks not only by its much larger mass, but also by its 
lifetime that is expected to be shorter than the QCD scale typical of
the formation of  hadronic bound states~\cite{pdg}. Within the SM, the 
top quark decays almost exclusively into a $W$ boson and a $b$ quark. The total decay width $\Gamma_t$ 
is therefore expected to be dominated by the partial decay width $\Gamma(t\to Wb)$. 
Neglecting higher order electroweak corrections and terms of order $m_b^2/m_t^2$, $\alpha_s^2$, 
and $(\alpha_s/\pi)M_W^2/m_t^2$, the partial width predicted by the SM at next-to-leading order (NLO) 
is~\cite{Jezabek:1989}
\begin{eqnarray}
\label{eq:smwidth}
\nonumber\Gamma (t\to Wb)_{\rm SM} &=& \frac{G_Fm_t^3}{8\pi\sqrt{2}}\, |V_{tb}|^2 \
\!\!\left ( 1 - \frac{M_W^2}{m_t^2} \right )^2 \!\!\left (1 +
2\frac{M_W^2}{m_t^2} \right ) \times \\
&& \!\! \times  \left [ 1 - \frac{2\alpha_s}{3\pi}  
\!\!\left ( \frac{2\pi^2}{3} - \frac{5}{2} \right )
\right ],
\end{eqnarray}
where $m_t$ ($m_b$) is the mass of the top (bottom) quark, $G_F$ ($\alpha_s$) the Fermi (strong interaction) 
coupling constant, $M_W$ the mass of the $W$ boson, and $V_{tb}$ the strength of the left-handed $Wtb$ coupling.
Setting $\alpha_s (M_Z)= 0.118$, $G_F = 1.16637\times 10^{-5}~{\rm GeV}^{-2}$, $M_W = 80.399$~GeV, $|V_{tb}|=1$~\cite{pdg}, 
and assuming $m_t=172.5$~GeV, we obtain ${\Gtbw_{\rm SM}}=1.33$~GeV.
A deviation from the theoretical prediction would indicate the presence of
 beyond SM (BSM) physics, including those involving BSM decays of the
 top quark to final states that escape detection. 
Examples of such BSM scenarios are the anomalous $Wtb$ couplings~\cite{Abazov:2008sz}, 
hadronically decaying charged Higgs bosons~\cite{mhdm} or a fourth generation $b'$ quark.

\begin{figure}
\vspace{0.1in}
\includegraphics[width=0.45\textwidth]{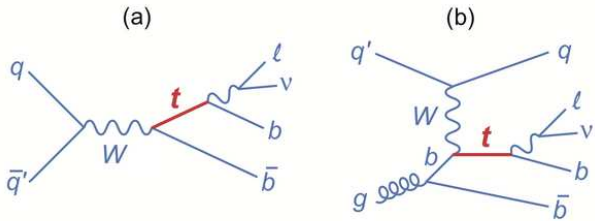}
\vspace{-0.1in}
\caption{Tree-level Feynman diagrams for (a) $s$ and (b) $t$-channel single top quark production.}
\label{fig:feynman}
\end{figure}

The electroweak single top quark production at the Tevatron proceeds
mainly through the exchange of a virtual $W$ boson accompanied at tree
level by a
$b$ quark in the $s$ channel~\cite{singletop-cortese}, or by both a
$b$ and a light quark in the $t$
channel~\cite{singletop-willenbrock,singletop-yuan}. A third channel,
$tW$, in which the top quark is produced in association with a $W$
boson, is not considered in this analysis because the expected
production cross section at the Tevatron is
small~\cite{singletop-xsec-kidonakis}.  Figure~\ref{fig:feynman} shows
the tree-level Feynman diagrams for $s$- and $t$-channel
production~\cite{charge-conjugate}. In this Letter, we determine
$\Gtbw$ from a measurement of the $t$-channel single top quark
production cross section, making use of the fact that the process
involves a $Wtb$ vertex and is thus proportional to $\Gtbw$. The
$t$-channel was chosen as it has the highest production cross section
at the Tevatron~\cite{singletop-xsec-kidonakis} and because BSM
contributions may have different effects on the $s$- and $t$-channel
cross sections. Here we do not assume that the $s$-channel production
rate is as predicted by the SM.  

The first direct upper bound on $\Gamma_t$~\cite{Aaltonen:2008ir} was set by the CDF Collaboration 
from an analysis of the invariant mass distribution of \ttbar\ candidate events using 1~fb$^{-1}$ of 
integrated luminosity. The first indirect determination of $\Gamma_t$~\cite{d0-topwidth-2.3} was obtained 
by the D0 Collaboration by combining the measurement of the single top $t$-channel cross section using 
2.3~fb$^{-1}$ of integrated luminosity~\cite{d0-tchannel-2.3} and the branching fraction $\Btbw$ determined 
from a sample of \ttbar\ events in 0.9~fb$^{-1}$ of integrated luminosity~\cite{d0-rb-0.9}. 
This method assumes the $W\to tb$ coupling in single top quark 
production is the same as in top quark decay. 

In this Letter, we apply the method of~\cite{d0-topwidth-2.3} in a new indirect determination 
of $\Gamma_t$ that is 
based on two prior D0 measurements, both performed using 
5.4~fb$^{-1}$ of integrated luminosity: 
the single top quark $t$-channel cross 
section $\sigma({\ppbar}{\rargap}tqb+X) = 2.90 \pm 0.59\;\rm (stat+syst)\; pb$~\cite{d0-tchannel-5.4} 
and the ratio
$R = \mathcal{B}(t \rightarrow Wb)/\mathcal{B}(t \rightarrow W q) =  0.90 \pm 0.04$~\cite{d0-rb-5.4}, where $q$ can
be a $d$, $s$ or $b$ quark.  

The partial decay width $\Gtbw \equiv \Gamma_p$ can be expressed in terms of the
$t$-channel single top quark production cross 
section as 
\begin{equation}
\label{eq:partwidth}
\Gtbw = \sigma(t\mathrm{-channel}) \ \frac{ \Gtbw_{\rm SM}}
{\sigma(t\mathrm{-channel})_{\rm SM}}.
\end{equation}
The total decay width $\Gamma_t$ can be written in terms of the partial decay width and the branching 
fraction $\Btbw$ as 
\begin{equation}
\label{eq:tot_width}
\Gamma_t = \frac{\Gamma_p}{\Btbw}. 
\end{equation}
Combining Eqs.~\ref{eq:tot_width} and~\ref{eq:partwidth}, the total decay width becomes
\begin{equation}
\label{eq:totwidth_derivation}
\Gamma_t = \frac{\sigma(t\mathrm{-channel}) \ \Gtbw_{\rm SM}}
  {\Btbw \ \sigma(t\mathrm{-channel})_{\rm SM}},
\end{equation}
from which it is possible to derive the lifetime of the top quark  as $\tau_{t}=\hbar/\Gamma_t$.

We can also use the measured value of $\Gamma_p$ to probe the $Wtb$ interaction and 
directly determine the Cabibbo-Kobayashi-Maskawa (CKM) quark mixing matrix~\cite{CKM} element $|V_{tb}|$. 
A direct determination of $|V_{tb}|$, without
assuming unitarity of the CKM matrix or three generations of quarks, is possible
through the measurement of the total single top quark
production cross section~\cite{singletop-vtb-jikia}. However, this method assumes that 
the top quark decays exclusively to $Wb$, and assumes the relative production rate of $s$ and $t$-channel 
single top production is as predicted by the SM. These two assumptions are removed when we combine 
the branching fraction measurement (which allows for $t\to Wd$ and $t\to Ws$ decays) and the single 
top cross section measured from the $t$-channel, independently of any assumption on the 
$s$-channel rate or on the ratio of $s$- to $t$-channel production cross sections.

\section{Analysis Method}
\label{sec:analysis}

This analysis relies on two prior D0 measurements, 
the single top $t$-channel cross section~\cite{d0-tchannel-5.4} and 
the ratio of the top quark branching fraction~\cite{d0-rb-5.4}. 
Both are based on 5.4~fb$^{-1}$ of integrated luminosity. The latter 
is performed by distinguishing between the standard decay mode of the top quark,
$t\bar{t} \rightarrow W^{+}bW^{-}\bar{b}$, (indicated by $bb$),
and decay modes that include light quarks ($q_l = d, s$),
$t\bar{t}\rightarrow W^{+}bW^{-}\bar{q}_l$ ($bq_l$) and
$t\bar{t} \rightarrow W^{+}q_lW^{-}\bar{q}_l$ ($q_lq_l$).
The analysis relies on a
sample of $ t\overline{t}$ events 
in which one $W$ boson decays into a quark and an antiquark and the other into 
an electron or muon and a neutrino, or events in which both $W$ bosons 
decay into $\ell \nu$. In both cases, we accept events in which the $W$ boson 
decayed to a $\tau$ lepton that subsequently decayed into an electron or a muon.
We use a neural network $b$-tagging algorithm~\cite{D0btag} to identify jets 
that originate from the hadronization of long-lived $b$~hadrons ($b$-tagged jet) 
and distinguish between the $bb$, $bq_l$ and $q_lq_l$ final states in $\ttbar$ decay.

The $t$-channel cross section measurement uses events containing an
isolated electron or muon, missing transverse energy and at least two
jets.  Background is suppressed by requiring that one or two of the
jets is identified as a $b$-jet.  The main background contributions
arise from $W$~bosons produced in association with jets and from {\ttbar}
pairs.  We further improve the discrimination between signal and
background by employing multivariate analysis techniques as described
in~\cite{d0-prd-2008}.  We use a discriminant trained to separate the
$t$-channel signal from the backgrounds in 6 independent analysis
channels, defined according to jet multiplicity (2, 3 or 4), and
number of $b$-tagged jets (1 or 2)~\cite{d0-tchannel-5.4}.

Based on the $t$-channel output discriminant distribution, we define a binned likelihood
\begin{equation}
L(\mathbf{D}|\mathbf{d}) = \prod^{M}_{i=1} \frac{e^{-d_i}d^{D_i}_i}{D_i!},
\end{equation}
where $\mathbf{D}$ and $\mathbf{d}$ are arrays containing the observed and mean expected count for all $M$
bins from the six different analysis channels. The mean expected count
can be written in terms of the partial ($\Gamma_p$) or total ($\Gamma_t$) top quark width as
\begin{equation}
\label{eq.meancount}
\mathbf{d}(\Gamma_{\{p,t\}}, \sigma_s', \mathbf{a}_t, \mathbf{a}_s, \mathbf{b}) =  c_{\{p,t\}} \Gamma_{\{p,t\}} \mathbf{a}_t + \sigma_s' \mathbf{a}_s + \mathbf{b}
\end{equation}
where $\sigma_s'$ is the $s$-channel cross section times $\mathcal{B}(t\rightarrow Wb)$, $\mathbf{a}_t$ and $\mathbf{a}_s$ are arrays containing 
the product of the acceptance and the integrated luminosity in each bin for $t$ and $s$ processes, respectively, 
and $\mathbf{b}$ is an array containing the mean count of expected background events. The term $c_{\{p,t\}}$ is 
given by 
\begin{equation}
c_p = \frac{\mathcal{B}(t\rightarrow Wb)\sigma(t\mathrm{-channel})_{SM}}{\Gamma(t\rightarrow Wb)_{SM}}
\end{equation}
or by
\begin{equation}
c_t = \frac{\mathcal{B}^2(t\rightarrow Wb)\sigma(t\mathrm{-channel})_{SM}}{\Gamma(t\rightarrow Wb)_{SM}},
\end{equation}
when measuring the partial or total top quark decay width, respectively. 
The  extra $\mathcal{B}(t\rightarrow Wb)$ term with respect to Eqs.~\ref{eq:partwidth} 
and~\ref{eq:totwidth_derivation} is needed to remove the assumption of 
$\mathcal{B}(t\rightarrow Wb) = 1$ used when generating the single top
quark and $\ttbar$ samples. 
We then form a Bayesian probability density for the partial or total
width by integrating the expression
\begin{widetext}
\begin{equation}
p(\Gamma_{\{p,t\}}|\mathbf{D}) = \frac{1}{\mathcal{N}}\int L(\mathbf{D}|\Gamma_{\{p,t\}}, \sigma_s', \mathbf{a}_t, \mathbf{a}_s, \mathbf{b}) \pi(\Gamma_{\{p,t\}}) \pi(\sigma_s') \pi(\mathbf{a}_t) \pi(\mathbf{a}_s) \pi(\mathbf{b}) d\sigma_s d\mathbf{a}_t d\mathbf{a}_{s} d\mathbf{b},
\end{equation}
\end{widetext}
where $\pi(*)$ represent our prior knowledge of the parameters $\sigma_s'$, $\mathbf{a}_t$, 
$\mathbf{a}_s$ and $\mathbf{b}$~\cite{d0-tchannel-5.4}. The normalization constant $\mathcal{N}$ ensures 
that $\int p(\Gamma_{\{p,t\}}|\mathbf{D}) d\Gamma_{\{p,t\}} = 1$. The integration is performed assuming a 
positive and uniform probability density for $\Gamma_{\{p,t\}}$ and for $\sigma_s$. The other priors quantify our knowledge of the systematic uncertainties 
for the values of $\mathbf{a}_t$, $\mathbf{a}_s$ and $\mathbf{b}$. Each independent systematic source 
is modeled with a Gaussian of mean zero and width set to one 
standard deviation of the corresponding uncertainty.

\section{Systematic Uncertainties}
\label{sec:systematics}

The main systematic uncertainties affect the $t$-channel output  discriminant 
as well as the measured branching fraction $\mathcal{B}(t\rightarrow Wb)$, and are summarized in 
Table~\ref{tab:sys}. Common systematics that affect both the discriminant and $\mathcal{B}(t\rightarrow Wb)$ 
are taken as 100\% correlated. 
\begin{table}[!h!tbp]
\centering
\setlength{\tabcolsep}{16pt} {
\begin{tabular}{lc}
\hline\hline
Sources  & Size [\%] \\
\hline
\multicolumn{2}{c}{\underline{Uncertainties on ${\cal B}(t\rightarrow Wb)$}}  \\
$b$-jet identification & 4.0 \\  
 $t\overline{t}$ cross section& 2.1 \\  
Integrated luminosity & 1.6 \\
Statistical uncertainty & 2.3 \\
Statistical correlation & 4.2 \\
\hline
\multicolumn{2}{c}{\underline{Uncertainties on $\sigma(t-\text{channel})$}} \\ 
$b$-jet identification & 9.3 \\  
$t\overline{t}$ cross section & 3.1  \\  
Integrated luminosity & 5.1 \\
$W$+jets normalization & 8.1 \\ 
Jet energy resolution & 11.6 \\
Jet energy scale & 6.8 \\
Monte Carlo statistics  & 6.7 \\  
\hline\hline
\end{tabular}}
\caption{Sources of statistical and main systematic uncertainties relative to the measured value for t-channel cross section and branching fraction that affects the determination of the partial/total decay width. We list the the most important uncertainties for the branching fraction and t-channel cross section measurements, respectively.}
\label{tab:sys}
\end{table}

The terms included in the uncertainty calculation are: 
\begin{itemize}
\item Uncertainty on jet flavor identification involving $b$, $c$ and light-flavor jet tagging rates and
the calorimeter response to $b$-jets.
\item Uncertainties on the modeling of the $\ttbar$ process used in the \Btbw\ measurement,
including uncertainties from parton distribution functions (PDFs), different event generators and hadronization models.
These uncertainties are correlated with the $\ttbar$ background yield uncertainty in the
$t$-channel discriminant.
\item Uncertainty on the integrated luminosity.
\item Uncertainties on the normalization of the $W$+jets heavy-flavor contribution.
\item Uncertainty in the jet energy scale and resolution.
\item The statistical uncertainty of the \Btbw\ measurement is added as systematic uncertainty. The statistical uncertainty related to the $t$-channel cross section is by construction included in the top quark width posterior distributions.
\item A systematic uncertainty is added to account for the statistical correlation 
due to a small overlap between events selected to build the $t$-channel
discriminant and those used in the \Btbw\ measurement.
\item Signal and background yield uncertainty because of the amount of
  Monte Carlo events used to
construct the background and signal discriminant.
\end{itemize}

Other terms included in the calculation of $t$-channel discriminant but not mentioned in the table are:
\begin{itemize}
\item Uncertainties on modeling the single top quark signal, including 
initial- and final-state radiation, scale uncertainties and PDFs.
\item Detector simulation uncertainty arising from the modeling of particle identification in the simulated
samples.
\item Uncertainties arising from the modeling of the different background sources that are obtained 
using data-driven methods. 
\end{itemize}

\section{Result}
\label{sec:result}

\begin{figure}[t]
\includegraphics[width=3.0in]{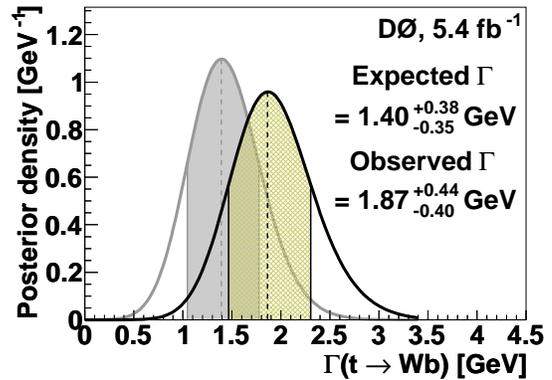}
\vspace{-0.5cm}
\caption{[color online] Probability density for the expected and measured partial width $\Gamma_p$. 
The shaded areas represent one standard deviation centered on the most probable value.} 
\label{fig:tqb_Gbtw_posterior}
\end{figure}

The expected and observed probability densities for the partial width $\Gamma_p$ are shown 
in Fig.~\ref{fig:tqb_Gbtw_posterior}.
The most probable value for the partial width is defined by the peak of the probability density 
function and corresponds to 
\begin{equation}
\label{eq:partialw_result}
\Gamma_p = 1.87^{+0.44}_{-0.40}~\mathrm{GeV}.
\end{equation}

\begin{figure}[!h!tbp]
\includegraphics[width=3.0in]{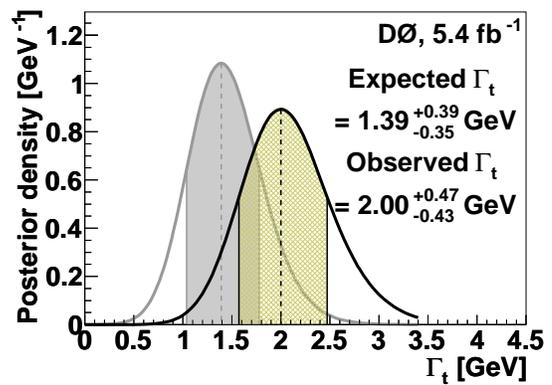}
\vspace{-0.5cm}
\caption{[color online] Probability density for the expected and measured total width $\Gamma_t$. The shaded areas represent one standard deviation around the most probable value.} 
\label{fig:total_Gbtw_posterior}
\end{figure}

The expected and observed probability densities for the total width 
$\Gamma_t$ are shown in Fig.~\ref{fig:total_Gbtw_posterior}. 
The total top quark width is found to be
\begin{equation}
\Gamma_t = 2.00^{+0.47}_{-0.43}~\mathrm{GeV},
\end{equation}
which can be expressed as a top quark lifetime of 
\begin{equation}
\tau_t = (3.29^{+0.90}_{-0.63}) \times10^{-25}~{\rm s}.
\end{equation}
This also translates in an upper limit to the top quark lifetime of 
$\tau_t < 4.88 \times 10^{-25}~\mathrm{s~at~95\%~C.L.}$.

\section {Top Quark Mass Dependence}

The measured branching fraction and $t$-channel production cross section, as well as 
$\Gamma(t \rightarrow Wb)_{\text{SM}}$, depend on the top quark mass $m_t$.
To study this dependency,  we repeat the analysis  using 
simulated {\ttbar} and single top samples that were generated at different values of $m_t$ in the 
range $170$ to $175\;\rm GeV$. The value of 
$\Gamma(t \rightarrow Wb)_{\text{SM}}$ is recalculated depending on $m_t$. 
Given that the dependence from $m_t$ is small, the value and
uncertainties for \Btbw\   corresponding to  $m_t = 172.5$~GeV are used in all cases.

Table~\ref{tab:topmass} summarizes the variation of the partial and total top quark decay width
as a function of $m_t$. The table also lists the values of $\Gamma(t \rightarrow Wb)_{\text{SM}}$ used 
in the analysis. The variation of the decay width with $m_t$ follows the 
non-monotonic variation observed for the $t$-channel cross section~\cite{d0-tchannel-5.4}.
\begin{table}
\begin{center}
\begin{tabular}{lccc}
\hline\hline
$m_t$ (GeV) & 170  & 172.5  & 175 \\
\hline
$\Gamma(t \rightarrow Wb)_{\text{SM}}$ (GeV) & $1.26$ & $1.33$ & $1.40$ \\
$\Gamma(t \rightarrow Wb)$ (GeV) & $1.48^{+0.36}_{-0.34}$ & $1.87^{+0.44}_{-0.40}$ & $1.76^{+0.40}_{-0.38}$ \\
$\Gamma_t$ (GeV) & $1.63^{+0.41}_{-0.38}$ & $2.00^{+0.47}_{-0.43}$ & $1.96^{+0.46}_{-0.43}$ \\
\hline\hline
\end{tabular}
\end{center}
\caption{Observed partial and total top quark decay width as a function of the top quark mass. 
We provide also the values for SM top quark partial widths used in the analysis.\label{tab:topmass}}
\end{table}

The effect of the mass dependency can be quantified by interpolating the observed $\Gamma_t$ in 
function of top mass from table~\ref{tab:topmass} to the current Tevatron combination
$m_t = 173.2 \pm 0.9$~GeV~\cite{tevatron-mass}. Adding a mass uncertainty of 
\[\delta\Gamma_t = \operatorname*{\max}_{m_t \in [172.5,175]}|\Gamma_t(m_t) - \Gamma_t(173.2)|\approx 0.07\text{ GeV}\]
in quadrature to the symmetrized interpolated error for $m_t = 173.2$~GeV results in a value for the total width of $\Gamma_t=2.03 \pm 0.46$~GeV,
and a value for top quark lifetime of $\tau_t=3.24_{-0.60}^{+0.95}\cdot10^{-25}$ s. A lower limit on the total decay width $\Gamma_{t} > 1.37$ GeV at
$95\%$ C.L. can be estimated by assuming that the posterior density for $\Gamma_t$ approximates 
a Gaussian distribution. This translates into an upper limit on the top quark lifetime of $\tau_{t} < 4.82 \cdot 10^{-25}$ s at $95\%$ C.L.
Therefore we conclude that the effect of the mass uncertainty is
 small with respect 
to the observed uncertainty obtained assuming a top quark mass of $m_t = 172.5$ GeV. 

\section {Measurement of \boldmath $|V_{tb}|$}

\begin{figure}
\vspace{0.1in}
\includegraphics[width=0.46\textwidth]{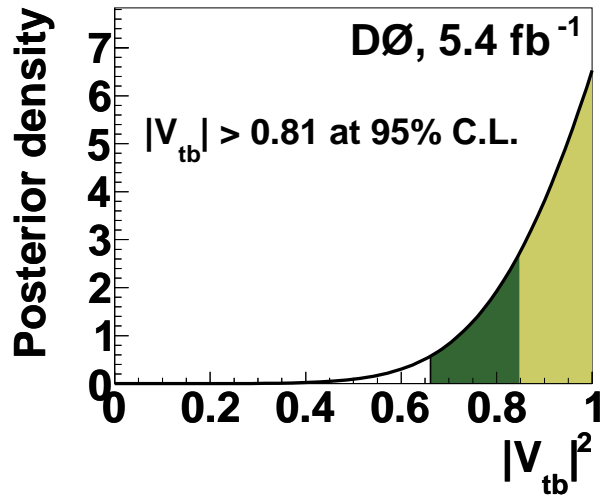}
\vspace{-0.1in}
\caption{[color online] Distributions of the posterior density for $|V_{tb}|^2$. The shaded (darked shaded) band indicates regions of 68\% (95\%) C.L.}
\label{fig:vtb}
\end{figure}

We construct a posterior probability density for $|V_{tb}|$ by setting 
\begin{equation}
c_{\{p,t\}} \Gamma_{\{p,t\}} = |V_{tb}|^2 \mathcal{B}(t\rightarrow Wb)\sigma(t\mathrm{-channel})_{SM, |V_{tb}|=1} 
\end{equation}
 in Eq.~\ref{eq.meancount}. A 
lower limit of $|V_{tb}| > 0.81$ at the 95\% C.L. is obtained by restricting 
the prior for $|V_{tb}|^2$ to be uniform for $0 \le |V_{tb}|^2 \le 1$, as illustrated in Fig.~\ref{fig:vtb}. 
A systematic uncertainty on the theoretical prediction for the $t$-channel cross section was included in the 
result. 

We apply the same procedure to constrain the strength of the coupling 
of a fourth generation $b'$ quark to the top quark $|V_{tb'}|$.
For this measurement we assume that $m_{b'}>m_t-m_W$, a small
probability density for the $b'$ quark to exist  in 
protons and antiprotons, 
and unitarity of the four-generation quark-mixing matrix with 
$|V_{tb}|^2+|V_{tb'}|^2=1$, and $|V_{td}|, |V_{ts}| \ll~1$. 
Using the limit on $|V_{tb}|$ and the condition $|V_{tb'}|^2 = 1 -  |V_{tb}|^2$,
we obtain  $|V_{tb'}| < 0.59$ at the 95\% C.L.

\section{Summary}

We have presented an improved determination of the width of the top quark using the Bayesian 
techniques previously used to measure the single top quark production cross section and an 
improved measurement of the branching fraction \Btbw. The method assumes that the coupling 
leading to $t$-channel 
single top quark production is identical to the coupling in the top quark decay. 
We have determined the top quark width as  
$\Gamma_t = 2.00^{+0.47}_{-0.43}~\mathrm{GeV}$ for $m_t=172.5$~GeV, which corresponds to a top quark 
lifetime of $\tau_t = (3.29^{+0.90}_{-0.63})\times10^{-25}~{\rm
  s}$. These are the most precise determinations of width and lifetime
to date. 
In addition, we set a lower limit of 
$|V_{tb}| > 0.81$ at the 95\% C.L. without assuming that the top quark decays exclusively to $Wb$ 
and with no assumption on the $s$- and $t$-channel relative production rate. We also set a limit on the 
strengths of the coupling for a fourth-generation $b'$ quark to the top quark of $|V_{tb'}| < 0.59$ at 95\% C.L.

%
We thank the staffs at Fermilab and collaborating institutions,
and acknowledge support from the
DOE and NSF (USA);
CEA and CNRS/IN2P3 (France);
FASI, Rosatom and RFBR (Russia);
CNPq, FAPERJ, FAPESP and FUNDUNESP (Brazil);
DAE and DST (India);
Colciencias (Colombia);
CONACyT (Mexico);
NRF (Korea);
CONICET and UBACyT (Argentina);
FOM (The Netherlands);
STFC and the Royal Society (United Kingdom);
MSMT and GACR (Czech Republic);
BMBF and DFG (Germany);
SFI (Ireland);
The Swedish Research Council (Sweden);
and
CAS and CNSF (China).
%



\begin{thebibliography}{99}

\bibitem{pdg}
  K.~Nakamura et al. (Particle Data Group)
  J.\ Phys\ G {\bf 37}, 075021 (2010).

\bibitem{Jezabek:1989}
  M.~Jezabek and J.~H.~K{\"u}hn,
  Nucl.\ Phys.\  B {\bf 314}, 1 (1989).

\bibitem{Abazov:2008sz}
  V.~M.~Abazov {\it et al.}  (D0 Collaboration),
  Phys.\ Rev.\ Lett.\  {\bf 101}, 221801 (2008).

\bibitem{mhdm} Y.~Grossman, Nucl.~Phys.~{\bf B426}, 355 (1994).

\bibitem{singletop-cortese}
  S.~Cortese and R.~Petronzio,
  Phys.\ Lett.\ B {\bf 253}, 494 (1991).

\bibitem{singletop-willenbrock}
  S.~S.~D.~Willenbrock and D.~A.~Dicus,
  Phys.\ Rev.\ D {\bf 34}, 155 (1986).

\bibitem{singletop-yuan}
  C.-P.~Yuan,
  Phys.\ Rev.\ D {\bf 41}, 42 (1990).

\bibitem{singletop-xsec-kidonakis}
  N.~Kidonakis,
  Phys.\ Rev.\ D {\bf 74}, 114012 (2006). The cross sections for the single top quark processes ($m_t = 172.5$~GeV) are $1.04 \pm 0.04$~pb ($s$ channel), $2.26 \pm 0.12$~pb ($t$ channel), and $0.28 \pm 0.06$~pb ($tW$ channel).

\bibitem{charge-conjugate}
  Unless otherwise stated, charge-conjugate states are implied.

\bibitem{Aaltonen:2008ir}
  T.~Aaltonen {\it et al.} (CDF Collaboration),
  Phys.\ Rev.\ Lett.\  {\bf 102}, 042001 (2009).

\bibitem{d0-topwidth-2.3}
  V.~M.~Abazov {\it et al.} (D0 Collaboration),
  Phys.\ Rev.\ Lett.\  {\bf 106}, 022001 (2011).

\bibitem{d0-tchannel-2.3}
  V.~M.~Abazov {\it et al.} (D0 Collaboration),
  Phys.\ Lett.\  B {\bf 682}, 363 (2010).

\bibitem{d0-rb-0.9}
  V.~M.~Abazov {\it et al.}  (D0 Collaboration),
  Phys.\ Rev.\ Lett.\  {\bf 100}, 192003 (2008).
  Although the original publication used $m_t=175$~GeV, 
  the \Btbw\ value used in this Letter is derived for $m_t$ = 170~GeV
  to be consistent with the $t$-channel cross section measurement.

\bibitem{d0-tchannel-5.4}
  V.~M.~Abazov {\it et al.} (D0 Collaboration),
  Phys.\ Lett.\ B {\bf 705}, 313 (2011)  

\bibitem{d0-rb-5.4}
  V.~M.~Abazov {\it et al.} (D0 Collaboration),  Phys. Rev. Lett. {\bf 107}, 121802 (2011).

\bibitem{CKM}
  N.~Cabibbo, Phys.\ Rev.\ Lett.\ {\bf 10}, 531 (1963);
  M.~Kobayashi and T.~Maskawa, Prog.\ Theor.\ Phys.\ {\bf 49}, 652 (1973).

\bibitem{singletop-vtb-jikia}
  G.~V.~Jikia and S.R.~Slabospitsky,
  Phys.\ Lett.\ B {\bf 295}, 136 (1992).

\bibitem{D0btag}
  V.~M.~Abazov {\it et al.} (D0 Collaboration),
  Nucl. Instrum. Methods in Phys. Res. Sect. A {\bf 620}, 490 (2010).

\bibitem{d0-prd-2008}
  V.~M.~Abazov {\it et al.} (D0 Collaboration),
  Phys.\ Rev.\ D\ {\bf 78}, 012005 (2008).

\bibitem{tevatron-mass}
The Tevatron Electroweak Working Group
for the CDF and D0 Collaborations,
 arXiv:1107.5255 [hep-ex].
 
\end{thebibliography}
\end{document}